\documentclass{Interspeech2024}

\interspeechcameraready 
\usepackage{multirow}
\usepackage{cite}
\usepackage{footmisc}
\makeatletter
\newcommand\footnoteref[1]{\protected@xdef\@thefnmark{\ref{#1}}\@footnotemark}
\makeatother

\title{ERes2NetV2: Boosting Short-Duration Speaker Verification Performance with Computational Efficiency}

\name[affiliation={1}]{Yafeng}{Chen}
\name[affiliation={1}]{Siqi}{Zheng}
\name[affiliation={1}]{Hui}{Wang}
\name[affiliation={1}]{Luyao}{Cheng}
\name[affiliation={1}]{Qian}{Chen}
\name[affiliation={1}]{Shiliang}{Zhang}
\name[affiliation={2}]{Junjie}{Li}

\address{
  $^1$Speech Lab, Alibaba Group, Hangzhou, China\\
  $^2$University of Science and Technology of China, Hefei, China}
\email{\{chenyafeng.cyf, zsq174630\}@alibaba-inc.com}


\begin{document}

\maketitle

\begin{abstract}
    Speaker verification systems experience significant performance degradation when tasked with short-duration trial recordings.  To address this challenge, a multi-scale feature fusion approach has been proposed to effectively capture speaker characteristics from short utterances. Constrained by the model's size, a robust backbone Enhanced Res2Net (ERes2Net) combining global and local feature fusion demonstrates sub-optimal performance in short-duration speaker verification. 
    To further improve the short-duration feature extraction capability of ERes2Net, we expand the channel dimension within each stage. However, this modification also increases the number of model parameters and computational complexity. To alleviate this problem, we propose an improved ERes2NetV2\footnote{Model is publicly available at \url{https://modelscope.cn/models/iic/speech_eres2netv2_sv_zh-cn_16k-common/summary}} by pruning redundant structures, ultimately reducing both the model parameters and its computational cost.
    A range of experiments conducted on the VoxCeleb datasets exhibits the superiority of ERes2NetV2, which achieves EER of \textbf{0.61\%} for the full-duration trial, \textbf{0.98\%} for the 3s-duration trial, and \textbf{1.48\%} for the 2s-duration trial on VoxCeleb1-O, respectively\footnote{Code is publicly available at \url{https://github.com/modelscope/3D-Speaker}}. 
\end{abstract}

\section{Introduction}
Speaker verification (SV) is the task of determining whether a given speech utterance belongs to a claimed speaker identity. With the great success of deep learning, SV systems~\cite{DBLP:conf/icassp/SnyderGSPK18, DBLP:conf/interspeech/YuL20, DBLP:conf/interspeech/DesplanquesTD20, CAM, DBLP:journals/corr/abs-2303-00332, DBLP:conf/interspeech/ChungHMLHCHJLH20, DBLP:conf/interspeech/ZhengLS20, DBLP:conf/slt/ZhouZW21, DBLP:conf/interspeech/LiuCWWHQ22, DBLP:conf/nips/0004LW023, DBLP:journals/taslp/LiuLWL24} have achieved remarkable progress in recent few years. Two prevalent architectures dominate neural network-based SV systems: the Time-Delay Neural Network (TDNN) and the two-dimensional Convolutional Neural Network (CNN). TDNN is characterized by the ability to efficiently model long temporal contexts between sequential data, which can be naturally applied to speech-related tasks. One of the most popular systems is x-vector~\cite{DBLP:conf/icassp/SnyderGSPK18}, which adopts TDNN as backbone. Subsequently, D-TDNN~\cite{DBLP:conf/interspeech/YuL20} is introduced to improve the system performance by adopting bottleneck layers and dense connectivity. ECAPA-TDNN~\cite{DBLP:conf/interspeech/DesplanquesTD20} unifies one-dimensional Res2Block with squeeze-excitation~\cite{DBLP:conf/cvpr/HuSS18} and expands the temporal context of each layer, achieving significant improvement. Furthermore, CAM \cite{CAM} and CAM++~\cite{DBLP:journals/corr/abs-2303-00332} use D-TDNN as backbone and adopts a multi-granularity pooling to capture contextual information at different levels with lower computational complexity. For CNN-based SV systems, residual networks~\cite{DBLP:conf/cvpr/HeZRS16}, originally developed for image recognition, have been adopted for speaker recognition~\cite{DBLP:conf/interspeech/ChungHMLHCHJLH20}. It uses a two-dimensional convolutional neural network in both time and frequency axes. 
In order to improve model's multi-scale representation ability, ~\cite{DBLP:journals/pami/GaoCZZYT21} propose Res2Net which increases the number of available receptive fields and used in SV~\cite{DBLP:conf/slt/ZhouZW21}. DF-ResNet~\cite{DBLP:conf/interspeech/LiuCWWHQ22}proposes the depth-first idea of CNN to achieve a better trade-off on performance and complexity.

Although the current SV models~\cite{DBLP:conf/interspeech/DesplanquesTD20, DBLP:journals/corr/abs-2303-00332, DBLP:conf/interspeech/ChungHMLHCHJLH20} exhibit superior performance on public datasets such as VoxCeleb~\cite{NagraniCZ17, ChungNZ18} and 3D-Speaker~\cite{DBLP:journals/corr/abs-2306-15354}, their efficacy notably diminishes when verifying short utterances. Therefore, multi-scale feature fusion has been introduced to enhance short-duration speaker verification, where it plays a pivotal role~\cite{DBLP:conf/interspeech/JungKCJK20, DBLP:conf/icassp/KimSHY22, DBLP:conf/slt/MunJHK22, DBLP:journals/corr/abs-2305-12838}. Jung et al.~\cite{DBLP:conf/interspeech/JungKCJK20} enhance speaker-discriminative information of features from multiple layers via a top-down pathway and lateral connections. Kim et al.~\cite{DBLP:conf/icassp/KimSHY22} present a deep layer aggregation structure and extends dynamic scaling polices for variable-duration utterances. Mun et al.~\cite{DBLP:conf/slt/MunJHK22} develop the multi-scale TDNN networks with selective kernel attention. Chen et al.~\cite{DBLP:journals/corr/abs-2305-12838} propose an Enhanced Res2Net architecture (ERes2Net) that incorporates a local and global feature fusion mechanism. The hierarchical attentive feature fusion architecture employed in ERes2Net excels in capturing short-term speaker characteristics. 

However, we find that the effectiveness of ERes2Net in extracting short-duration features is limited by the model size. In addition, the global feature fusion in ERes2Net which modulates features of different temporal scales in bottom-up pathway possesses a certain degree of redundancy. Based on the above two points, we introduce ERes2NetV2, an improved version of ERes2Net, which integrates Bottom-up Dual-stage Feature Fusion (BDFF) and Bottleneck-like Local Feature Fusion (BLFF). BDFF integrates multi-scale feature maps between stages 3 and 4 within the bottom-up pathway to capture global information and reduce structural redundancy. Concurrently, BLFF expands the channel dimensions of the feature maps and subsequently compresses the channel dimensions of the segmented features inspired by bottleneck feature structure, with the intent of strengthening short-duration feature extraction and diminishing both the model parameters and computational complexity.

Experiments conducted on the public VoxCeleb and 3D-Speaker datasets, consistently demonstrate the superiority of our proposed approaches over baseline systems across different durations. The remainder of this paper is organized as follows. In Section 2, we elaborate on the modifications to the ERes2Net that are aimed at enhancing the robustness of short-duration feature extraction while simultaneously reducing computational complexity. The experimental setup, the results and analysis are presented in Section 3. Finally, conclusions are given in Section 4.

\section{ERes2NetV2 for speaker verification}

\subsection{Overview of ERes2NetV2}

Effective multi-scale feature fusion is essential for enhancing short-duration speaker verification performance. We use ERes2Net as a starting point to further strengthen its short-duration feature extraction capability. The proposed ERes2NetV2 improves the robustness of speaker embedding by expanding the channel dimension of features and pruning the redundant structures as shown in Fig. 1. It consists of two branches: a bottom-up dual-stage feature fusion branch and a bottleneck-like local feature fusion branch.

\vspace{-0.2cm}
\begin{figure}[htb]
  \centering
  \includegraphics[scale=0.7]{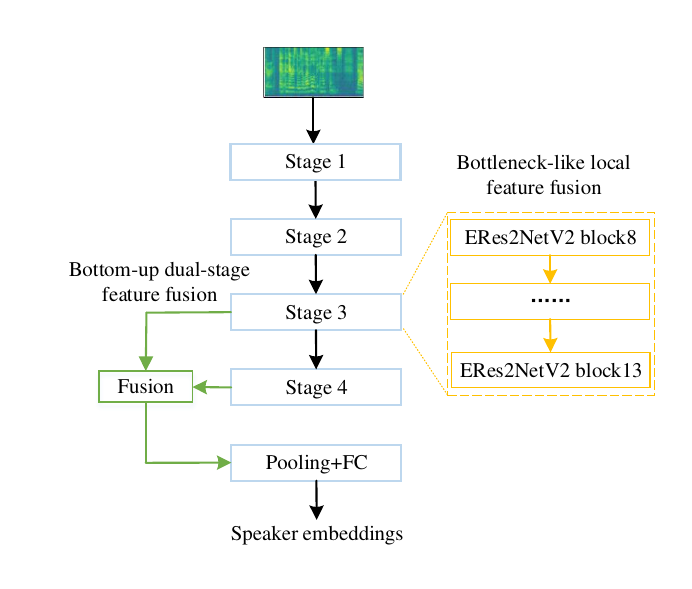}
  \vspace{-0.4cm}
  \caption{Overview of the ERes2NetV2 framework.}
\end{figure}

The BDFF processes acoustic features from a global perspective to aggregate signal. We optimize the original global feature fusion by removing the global feature fusion module interconnecting stages 1, 2, and 3. It preserves the validity of verification performance while significantly reducing the amount of parameters and calculation complexity.
The BLFF fuses the features within each residual block to extract the local signal. To augment the model's capacity for extracting speaker embeddings, we expand the channel dimensions of features in each residual block. Concurrently, we narrow the width in the split features to curtail the number of parameter and computational intricacy. 

\subsection{Bottom-up dual-stage feature fusion}

This section describes the BDFF component of ERes2NetV2, as shown in the green part of Fig. 1. BDFF aims to enhance the global feature interaction by modulating the output features of stage 3 and stage 4 in bottom-up pathway. At the stage 3 of ERes2NetV2, the network possess a larger receptive field compared to earlier stages, enabling the capture of more extensive contextual information. Features from stage 4 contain high-level speaker information. When fused with features from stage 3, the combined features complement the potential loss of fine details from stage 4, while enhancing the robustness of representations. Additionally, the fusion of features from stage 3 and stage 4 provides a balanced combination of medium to high-level features, ensuring that both detail and contextual information are adequately represented.

Specifically, we extract multi-scale features $\{\mathbf{S}_j | j = 3, 4\}$ from the final layer of ERes2NetV2 stage, which contain temporal information at varying resolutions. We then apply down-sampling to the feature maps from the output of ERes2NetV2 stage 3 along both time and frequency dimensions, utilizing a $3\times3$ convolutional kernel, while simultaneously doubling the channel dimension. Next, we modulate the set $\{\mathbf{S}_j | j = 3, 4\}$ by employing the Attentional Feature Fusion (AFF) module. This module computes attention weights with a global perspective. The down-sampled feature maps are enhanced with features via AFF module as follows:
\begin{equation}
\mathbf{F} = AFF[D(\mathbf{S}_{j-1}), \mathbf{S}_j] \qquad j = 4
\end{equation}
where $D(\cdot)$ denotes the down-sampling operation. $\mathbf{F}$ stands for the fusion of the $(j-1)$th stage output and the $j$th stage output in the bottom-up pathway. 
$AFF$ module takes the concatenation of adjacent feature maps $\mathbf{x}$ and $\mathbf{y}$ as the input. Then calculate the local attention weights $\mathbf{Att}$ as follows.
\begin{equation}
\mathbf{Att} = tanh(BN(\mathbf{W}_2 \cdot SiLU (BN(\mathbf{W}_1 \cdot [\mathbf{x}, \mathbf{y}]))))
\end{equation}
where $[\cdot]$ denotes the concatenation along the channel dimension. $\mathbf{W}_1$ and $\mathbf{W}_2$ are point-wise convolution with output channel sizes of $C/r$ and $C$ respectively. $r$ is the channel reduction ratio. $BN$ refers to batch normalization~\cite{DBLP:conf/icml/IoffeS15}. $SiLU(\cdot)$ and $tanh(\cdot)$ stand for Sigmoid Linear Unit (SiLU) and tanh activation function respectively. 

\subsection{Bottleneck-like local feature fusion}

This section describes the structure of BLFF in ERes2NetV2 block as shown in Fig. 2. Inspired by the bottleneck feature, we double the channel dimension of the input feature in each block, and then reduce the channel dimension using a 1×1 convolution compared with the ERes2Net block. Feature maps are split and concatenated through 3$\times$3 convolution kernels and an AFF module. These feature maps are then processed by a 1$\times$1 convolutional kernel to expand the channel dimension to match that of the input. The structure reduce the redundancy via expand-reduce-expand operator, aiming for a more efficient model with less computational cost. 

\vspace{-0.2cm}
\begin{figure}[htb]
  \centering
  \includegraphics[scale=0.68]{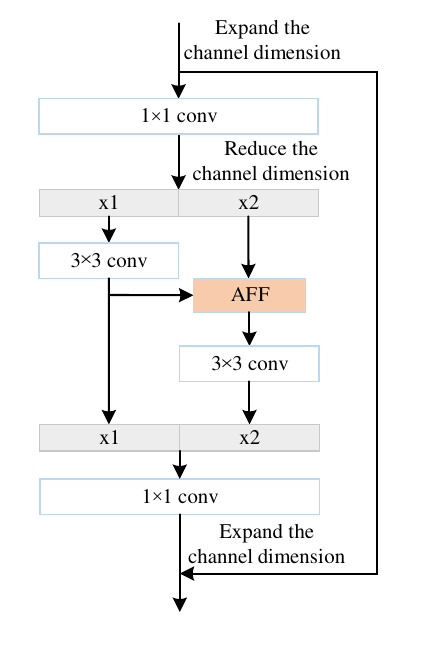}
  \vspace{-0.4cm}
  \caption{Illustration of bottleneck-like local feature fusion in ERes2NetV2 block.}
\end{figure}

\begin{table*}[thb]
    \small 
    \setlength{\tabcolsep}{2pt} 
    \caption{Results on VoxCeleb1-O, VoxCeleb1-E, VoxCeleb1-H and 3D-Speaker datasets. ERes2Net* refers to our replication of the baseline ERes2Net framework. ERes2NetV2 w/o BL denotes removing Bottleneck-like Local feature fusion on the basis of ERes2NetV2. ERes2NetV2 w/o (BL+BD) denotes removing Bottleneck-like Local feature fusion and Bottom-up Dual-stage feature fusion of ERes2NetV2, only expanding the channel dimensions within each stage of ERes2Net. We compare the number of parameters (Params) and floating-point operations (FLOPs) of different models. The best results for each test set are in bold.}
    \label{tab:main_results}
    \centering
    \begin{tabular}{c c c c c c c c c c c}
    \toprule
    \multirow{2}{*}{\textbf{Architecture}} & \multirow{2}{*}{\textbf{Params}} & \multirow{2}{*}{\textbf{FLOPs}} & \multicolumn{2}{c}{\textbf{VoxCeleb1-O}} & \multicolumn{2}{c}{\textbf{VoxCeleb1-E}} & \multicolumn{2}{c}{\textbf{VoxCeleb1-H}} & \multicolumn{2}{c} 
    {\textbf{3D-Speaker}}\\
    \cmidrule(lr){4-5} \cmidrule(lr){6-7} \cmidrule(lr){8-9} \cmidrule(lr){10-11}
    & & & \textbf{EER (\%)} & \textbf{MinDCF} & \textbf{EER (\%)} & \textbf{MinDCF} & \textbf{EER (\%)} & \textbf{MinDCF} & \textbf{EER (\%)} & \textbf{MinDCF}\\
    \midrule
    ERes2Net* & 6.61M & 5.16G & 0.84 & 0.088 & 0.96 & 0.102 & 1.78 & 0.175 & 7.12 & 0.657 \\
    \midrule
    ERes2NetV2 & 17.8M & 12.6G & 0.61 & \textbf{0.054} & 0.76 & 0.082 & 1.45 & \textbf{0.143} & 6.52 & \textbf{0.589} \\
    ERes2NetV2 w/o BL & 20.7M & 16.2G & 0.59 & 0.058 & 0.76 & 0.086 & 1.46 & 0.146 & 6.53 & 0.604\\
    ERes2NetV2 w/o (BL+BD) & 22.4M & 20.4G & \textbf{0.52} & 0.055 & \textbf{0.75} & \textbf{0.079} & \textbf{1.44} & 0.145 & \textbf{6.44} & 0.599 \\
    
    \bottomrule
    \end{tabular}
\end{table*}

\begin{table}[h]
    \caption{Tasks on VoxCeleb1 dataset. Here ‘O’ denotes ‘original’, ‘E’ denotes ‘extended’, and ‘H’ denotes ‘hard’}
    \scalebox{0.95}{
    \centering
        \begin{tabular}{cccccc}
        \toprule
         & VoxCeleb1-O & VoxCeleb1-E & VoxCeleb1-H \\
        \midrule
        Speakers & 40 & 1251 & 1251 \\
        Trials & 37,611 & 579,818 & 550,894 \\
        \bottomrule
        \end{tabular}}
\end{table}

\section{Experiments and analysis}

\subsection{Datasets and evaluation metrics}

We conduct experiments on the public speaker verification datasets, VoxCeleb~\cite{NagraniCZ17, ChungNZ18} and 3D-Speaker~\cite{DBLP:journals/corr/abs-2306-15354}, to evaluate the effectiveness of the proposed methods. For the VoxCeleb dataset, we utilize the development set of VoxCeleb2 for training, which contains 5,994 speakers. The evaluation is performed on three test trials: VoxCeleb1-O, VoxCeleb1-E, and VoxCeleb1-H, as detailed in Table 2. For the 3D-Speaker dataset, the training set comprises 10,000 speakers with a total of 579,013 utterances. The accumulated duration of valid speech amounts to 1,124 hours. The evaluation set uses multi-device dataset. The results are reported in terms of two metrics, namely, the equal error rate (EER) and the minimum of the normalized detection cost function (MinDCF) with the settings of $P_{target}$ = 0.01 and $C_{fa} = C_{miss}$ = 1.

\subsection{Data Augmentation} 
Due to the background noise, reverberation and laughter contained in the speech data, three data augmentation techniques are applied to improve the robustness of the system: online data augmentation \cite{DBLP:journals/taslp/CaiCZL20} with MUSAN corpus~\cite{DBLP:journals/corr/SnyderCP15} with SNR between 0 to 15 for additive noise, RIR dataset \cite{DBLP:conf/icassp/KoPPSK17} for reverberation, and speed perturb \cite{wang2020dku} with 0.9 and 1.1 times speed changes to treble the number of speakers.

\subsection{Implementation Details}
The acoustic features used in the experiments are 80-dimensional Filter Bank (FBank) with 25ms windows and 10ms shift. Speech Activity Detection (SAD) is not performed since the training data mostly consists of continuous speech. We use the stochastic gradient descent (SGD) optimizer with a cosine annealing scheduler and a linear warm-up scheduler. During the first 5 epochs, the learning rate is linearly increased to 0.2. The momentum is set to 0.9 and weight decay to 1e-4. Angular additive margin softmax (AAM-Softmax) loss is used for all experiments. The margin and scaling factors of AAM-Softmax loss are set to 0.3 and 32 respectively. We adopt the large margin fine-tuning \cite{DBLP:conf/icassp/ThienpondtDD21} strategy. The speaker embeddings are extracted from the first fully connected layer with a dimension of 192. 3-second segments are randomly cropped from each audio to construct the training mini-batches. Cosine similarity scoring is used for evaluation, without applying score normalization in the back-end.

\subsection{Analysis of performance and complexity}

We investigate the performance of proposed methods and evaluate them on the VoxCeleb1-O, VoxCeleb1-E, VoxCeleb1-H and 3D-Speaker trials. The experimental results are shown in Table 1. Comparing row 1 and 2 shows that ERes2NetV2 outperforms ERes2Net substantially and consistently across four test sets, achieving EERs of \textbf{0.61\%}, \textbf{0.76\%}, \textbf{1.45\%} and \textbf{6.52\%}, which demonstrates the solid verification performance in supervised SV. It yields relative improvements in EERs on the four test sets by \textbf{27.4\%}, \textbf{20.8\%}, \textbf{18.5\%} and \textbf{8.4\%}, respectively. The improved consistency observed on the four test sets confirms the robustness of the proposed method.

Next, we remove individual components to explore the contribution of each to the performance improvements. Comparing rows 2 and 3, it can be observed that bottleneck-like feature fusion achieves a relative reduction in parameters and computational complexity by \textbf{14.0\%} and \textbf{22.2\%}, respectively, without compromising on verification performance. 

Furthermore, comparing rows 3 and 4 demonstrates that global feature fusion among the initial three stages in ERes2Net is not essential for extracting discriminative speaker embeddings. By streamlining the structure and reducing redundancy, the bottom-up dual-stage feature fusion reduces the number of parameters and the computational complexity, achieving relative reductions of \textbf{7.6\%} and \textbf{20.6\%}, respectively.

\begin{table*}[t]
    \caption{Results on short-duration VoxCeleb1-O, VoxCeleb1-E, and VoxCeleb1-H datasets. All models in the table are reproduced through \textbf{3D-Speaker toolkit}\cite{chen20243d}.}
    \label{tab:main_results}
    \centering
    \begin{tabular}{cccccccccccccccccccccc}
    \toprule
    \multicolumn{1}{c}{\multirow{1}{*}{\textbf{}}}
    & \multicolumn{1}{c}{\multirow{1}{*}{\textbf{}}}
    & \multicolumn{4}{c}{\multirow{1}{*}{\textbf{VoxCeleb1-O}}}
    & \multicolumn{4}{c}{\multirow{1}{*}{\textbf{VoxCeleb1-E}}}
    & \multicolumn{4}{c}{\multirow{1}{*}{\textbf{VoxCeleb1-H}}} \\
    \cmidrule(lr){3-6} \cmidrule(lr){7-10} \cmidrule(lr){11-14}
    \multicolumn{1}{c}{\multirow{1}{*}{\textbf{Model}}} &
    & \multicolumn{2}{c}{\multirow{1}{*}{\textbf{EER(\%)}}}
    & \multicolumn{2}{c}{\multirow{1}{*}{\textbf{MinDCF}}}
    & \multicolumn{2}{c}{\multirow{1}{*}{\textbf{EER(\%)}}}
    & \multicolumn{2}{c}{\multirow{1}{*}{\textbf{MinDCF}}}
    & \multicolumn{2}{c}{\multirow{1}{*}{\textbf{EER(\%)}}}
    & \multicolumn{2}{c}{\multirow{1}{*}{\textbf{MinDCF}}} \\
    \cmidrule(lr){3-4} \cmidrule(lr){5-6} \cmidrule(lr){7-8} \cmidrule(lr){9-10} \cmidrule(lr){11-12} \cmidrule(lr){13-14}
    & & \textbf{3.0s}   & \multicolumn{1}{c}{\textbf{2.0s}}   & \textbf{3.0s}   & \multicolumn{1}{c}{\textbf{2.0s}}  & \textbf{3.0s} &{\textbf{2.0s}}  & \textbf{3.0s}   &{\textbf{2.0s}}  & \textbf{3.0s}   &{\textbf{2.0s}}   &  \textbf{3.0s}   & \textbf{2.0s} \\ 
    \midrule
    Res2Net & & 1.98 & 2.73 & 0.201 & 0.294 & 1.91 & 2.70 & 0.211 & 0.284 & 3.30 & 4.59 & 0.302 & 0.403 \\
    MFA-Conformer & & 1.84 & 2.71 & 0.216 & 0.301 & 1.92 & 2.83 & 0.215 & 0.298 & 3.43 & 4.83 & 0.318 & 0.422 \\
    ResNet34 & & 1.47 & 2.12 & 0.161 & 0.225 & 1.51 & 2.18 & 0.170 & 0.239 & 2.68 & 3.78 & 0.267 & 0.352 \\
    ECAPA-TDNN & & 1.27 & 1.95 & 0.172 & 0.246 & 1.33 & 1.97 & 0.153 & 0.224 & 2.59 & 3.71 & 0.261 & 0.354 \\
    ERes2Net & & 1.89 & 3.28 & 0.220 & 0.325 & 1.87 & 3.41 & 0.210 & 0.368 & 3.42 & 5.74 & 0.319 & 0.524 \\
    \midrule
    ERes2NetV2 & & 0.98 & 1.48 & 0.106 & 0.183 & 1.09 & 1.65 & 0.124 & 0.179 & 2.06 & 2.98 & 0.208 & 0.295 \\
    ERes2NetV2 w/o BL & & 0.99 & 1.52 & 0.096 & 0.174 & 1.09 & 1.64 & 0.127 & 0.180 & 2.07 & 3.04  & 0.215 & 0.299\\
    ERes2NetV2 w/o (BL+BD) & & 0.94 & 1.43 & 0.093 & 0.176 & 1.05 & 1.60 & 0.119 & 0.178 & 2.01 & 2.96 & 0.205 & 0.287\\

    \bottomrule
    \end{tabular}
\end{table*}

\begin{figure}[thb]
  \centering
  \includegraphics[scale=0.20]{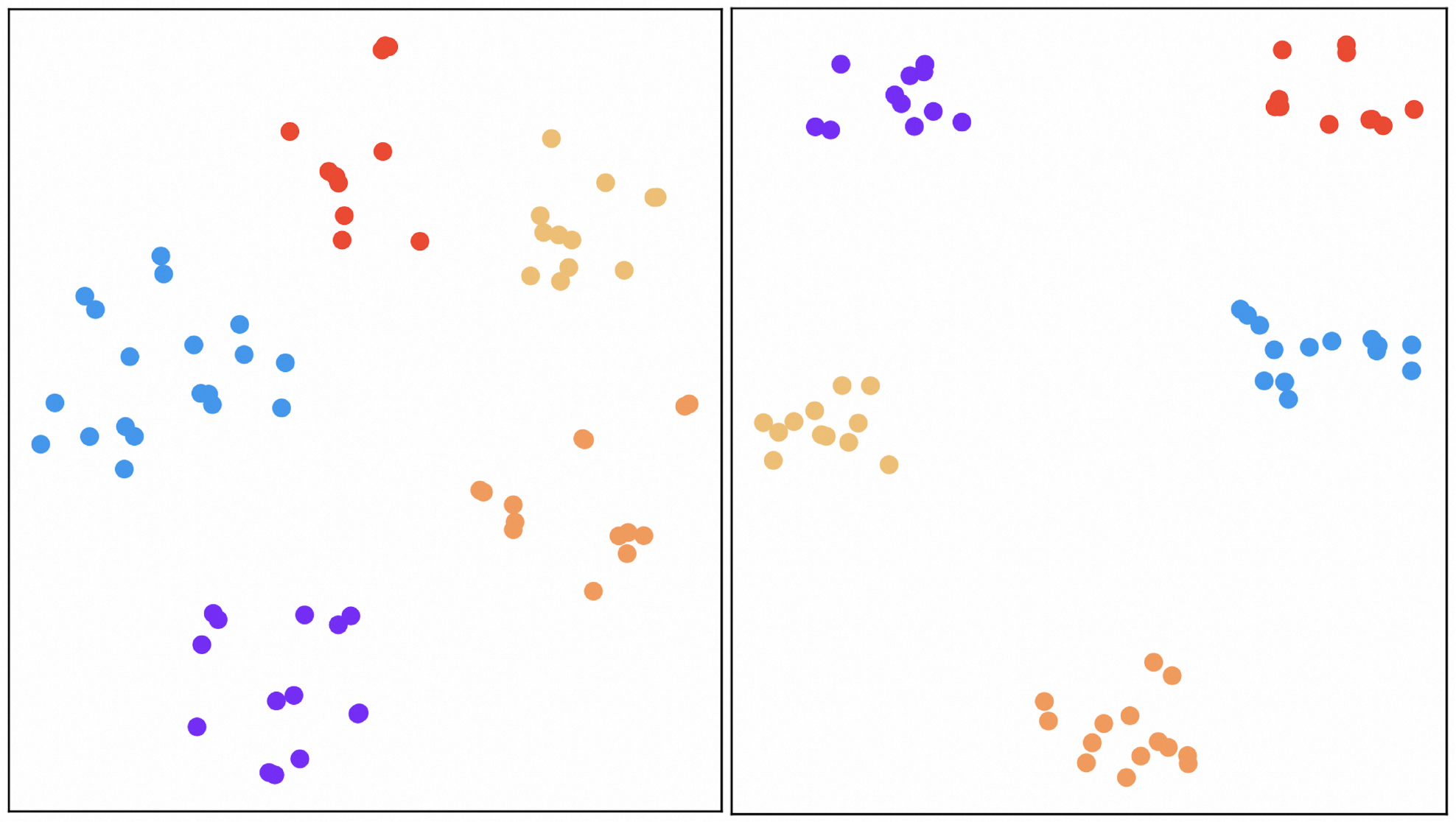}
  \caption{The t-SNE visualizations depict the extracted embeddings for five speakers, each denoted by a distinct color. The left figure displays the speaker embeddings derived from the baseline ERes2Net, while the right figure shows those obtained from our ERes2NetV2. The embeddings from ERes2NetV2 clearly exhibit enhanced separation compared to those from ERes2Net, suggesting improved discriminability.}
  \label{fig:visualization}
\end{figure}

\subsection{Comparison with short-duration utterances}

To evaluate the performance with short utterances on the VoxCeleb1-O, VoxCeleb1-E and VoxCeleb1-H trials, we use full-duration enroll utterances and test utterances truncated to durations of 2 and 3 seconds. The test utterance is cropped randomly in the utterance, and if the utterance length was shorter than the target length, it will be duplicated.

In Table 3, we compare ERes2NetV2 with the classic systems: Res2Net~\cite{DBLP:conf/slt/ZhouZW21}, MFA-Conformer~\cite{DBLP:conf/interspeech/ZhangLWZH0LM22}, ResNet34~\cite{DBLP:conf/interspeech/ChungHMLHCHJLH20}, ECAPA-TDNN~\cite{DBLP:conf/interspeech/DesplanquesTD20} and ERes2Net~\cite{DBLP:journals/corr/abs-2305-12838} for short-duration speaker verification. We observe that ECAPA-TDNN achieves competitive results in short-duration speaker verification. And the ERes2NetV2 achieves the impressive performance on the 3-second and 2-second test sets. It yields relative improvements in EERs on the three test sets by \textbf{48.1\%}, \textbf{41.7\%}, and \textbf{39.7\%} in 3-second trials condition and \textbf{54.8\%}, \textbf{51.6\%} and \textbf{48.1\%} in 2-second trials condition compared with ERes2Net. Significantly, the ERes2NetV2 demonstrates a substantial enhancement over ERes2Net, both in the absolute values and in the relative percentage reductions of EER and MinDCF. Comparing rows 6, 7, and 8 reveals that both BDFF and BLFF reduce the number of parameters and computational complexity, while resulting in only a minor decrease in verification performance for short-duration verification performance.

We employ the t-distributed Stochastic Neighbor Embedding (t-SNE) \cite{van2008visualizing} to compare the disentanglement performance of 2-second speaker embeddings derived from both ERes2Net and ERes2NetV2, as illustrated in Fig. \ref{fig:visualization}. It is clear that the embeddings extracted via ERes2NetV2 exhibit superior clustering capabilities compared to those from ERes2Net, suggesting that ERes2NetV2 makes short-duration speaker embeddings more discriminative.

\begin{table}[h]
    \caption{Comparison with published systems in VoxCeleb1-O, * means that data augmentation is not used. The MFA-Conformer model was trained utilizing the development sets of both VoxCeleb1 and VoxCeleb2, whereas the remaining models were trained on the development set of VoxCeleb1.}
    \centering
    \setlength{\tabcolsep}{4pt} 
    \begin{tabular}{l c c c}
    \toprule
    \textbf{Framework} & \textbf{Params (M)} & \textbf{EER (\%)} & \textbf{MinDCF} \\
    \midrule
    ResNet34-FPM* \cite{DBLP:conf/interspeech/JungKCJK20} & 5.85 & 1.98 & 0.205 \\
    Res2Net-MWA \cite{DBLP:journals/taslp/GuGZ23} & 5.50 & 1.71 & 0.165 \\
    Res2Net-26w8s \cite{DBLP:conf/slt/ZhouZW21} & 9.30 & 1.45 & 0.147 \\
    MFA-TDNN \cite{DBLP:conf/icassp/LiuDLL22} & 7.32 & 0.85 & 0.092 \\
    ERes2Net \cite{DBLP:journals/corr/abs-2305-12838} & 6.61 & 0.83 & 0.072 \\
    ECAPA-TDNN \cite{DBLP:conf/interspeech/ZhangLWZH0LM22} & 20.8 & 0.82 & 0.112 \\
    SKA-TDNN \cite{DBLP:conf/slt/MunJHK22} & 34.9 & 0.78 & 0.047 \\
    MFA-Conformer \cite{DBLP:conf/interspeech/ZhangLWZH0LM22} & 20.5 & 0.64 & 0.081 \\
    \midrule
    ERes2NetV2 \ (Ours) & 17.8 & 0.61 & 0.054 \\
    \bottomrule
    \end{tabular}
\end{table}

\subsection{Comparison with published systems}
The proposed ERes2NetV2 is compared with several state-of-the-art models. To ensure a fair comparison, we keep the training datasets the same with the reported systems in full-duration trials. As presented in Table 4, we compare ERes2NetV2 with other models such as ECAPA-TDNN, SKA-TDNN, MFA-Conformer and so on. Experimental results demonstrate that our model achieves better performance with fewer parameters, indicating the effectiveness and efficiency of our proposed approach for aggregating multi-scale features in time, frequency, and channel dimensions. 

\section{Conclusions}

In this paper, we propose ERes2NetV2, an improved variant of the ERes2Net, specifically designed to effectively capture features from short-duration utterances. ERes2NetV2 contains two key components: bottom-up dual-stage feature fusion and bottleneck-like local feature fusion. It exhibits an augmented capability for short-duration feature extraction. It culminates in an optimal equilibrium between computational efficiency and model performance. A series of experiments performed on the VoxCeleb datasets reveal the superior performance in short-duration speaker verification. For future work, we are interested in studying its performance in self-supervised speaker verification \cite{moco,rdino} and its potential for scaling.

\bibliographystyle{IEEEtran}
\bibliography{mybib}

\begin{thebibliography}{10}
\providecommand{\url}[1]{#1}
\csname url@samestyle\endcsname
\providecommand{\newblock}{\relax}
\providecommand{\bibinfo}[2]{#2}
\providecommand{\BIBentrySTDinterwordspacing}{\spaceskip=0pt\relax}
\providecommand{\BIBentryALTinterwordstretchfactor}{4}
\providecommand{\BIBentryALTinterwordspacing}{\spaceskip=\fontdimen2\font plus
\BIBentryALTinterwordstretchfactor\fontdimen3\font minus \fontdimen4\font\relax}
\providecommand{\BIBforeignlanguage}[2]{{%
\expandafter\ifx\csname l@#1\endcsname\relax
\typeout{** WARNING: IEEEtran.bst: No hyphenation pattern has been}%
\typeout{** loaded for the language `#1'. Using the pattern for}%
\typeout{** the default language instead.}%
\else
\language=\csname l@#1\endcsname
\fi
#2}}
\providecommand{\BIBdecl}{\relax}
\BIBdecl

\bibitem{DBLP:conf/icassp/SnyderGSPK18}
D.~Snyder, D.~Garcia{-}Romero, G.~Sell, D.~Povey, and S.~Khudanpur, ``X-vectors: Robust {DNN} embeddings for speaker recognition,'' in \emph{{ICASSP} 2018, Calgary, AB, Canada, April 15-20, 2018}.\hskip 1em plus 0.5em minus 0.4em\relax {IEEE}, 2018, pp. 5329--5333.

\bibitem{DBLP:conf/interspeech/YuL20}
Y.~Yu and W.~Li, ``Densely connected time delay neural network for speaker verification,'' in \emph{Interspeech 2020, Shanghai, China, 25-29 October 2020}.\hskip 1em plus 0.5em minus 0.4em\relax {ISCA}, 2020, pp. 921--925.

\bibitem{DBLP:conf/interspeech/DesplanquesTD20}
B.~Desplanques, J.~Thienpondt, and K.~Demuynck, ``{ECAPA-TDNN:} emphasized channel attention, propagation and aggregation in {TDNN} based speaker verification,'' in \emph{Interspeech 2020, Shanghai, China, 25-29 October 2020}.\hskip 1em plus 0.5em minus 0.4em\relax {ISCA}, 2020, pp. 3830--3834.

\bibitem{CAM}
Y.-Q. Yu, S.~Zheng, H.~Suo, Y.~Lei, and W.-J. Li, ``Cam: Context-aware masking for robust speaker verification,'' in \emph{ICASSP 2021 - 2021 IEEE International Conference on Acoustics, Speech and Signal Processing (ICASSP)}, 2021, pp. 6703--6707.

\bibitem{DBLP:journals/corr/abs-2303-00332}
\BIBentryALTinterwordspacing
H.~Wang, S.~Zheng, Y.~Chen, L.~Cheng, and Q.~Chen, ``{CAM++:} {A} fast and efficient network for speaker verification using context-aware masking,'' \emph{CoRR}, vol. abs/2303.00332, 2023. [Online]. Available: \url{https://doi.org/10.48550/arXiv.2303.00332}
\BIBentrySTDinterwordspacing

\bibitem{DBLP:conf/interspeech/ChungHMLHCHJLH20}
J.~S. Chung, J.~Huh, S.~Mun, and et~al., ``In defence of metric learning for speaker recognition,'' in \emph{Interspeech 2020, Shanghai, China, 25-29 October 2020}.\hskip 1em plus 0.5em minus 0.4em\relax {ISCA}, 2020, pp. 2977--2981.

\bibitem{DBLP:conf/interspeech/ZhengLS20}
S.~Zheng, Y.~Lei, and H.~Suo, ``Phonetically-aware coupled network for short duration text-independent speaker verification,'' in \emph{Interspeech 2020, 21st Annual Conference of the International Speech Communication Association, Virtual Event, Shanghai, China, 25-29 October 2020}.\hskip 1em plus 0.5em minus 0.4em\relax {ISCA}, 2020, pp. 926--930.

\bibitem{DBLP:conf/slt/ZhouZW21}
T.~Zhou, Y.~Zhao, and J.~Wu, ``Resnext and res2net structures for speaker verification,'' in \emph{{IEEE} Spoken Language Technology Workshop, {SLT} 2021}.\hskip 1em plus 0.5em minus 0.4em\relax {IEEE}, 2021, pp. 301--307.

\bibitem{DBLP:conf/interspeech/LiuCWWHQ22}
B.~Liu, Z.~Chen, S.~Wang, H.~Wang, B.~Han, and Y.~Qian, ``Df-resnet: Boosting speaker verification performance with depth-first design,'' in \emph{Interspeech 2022}.\hskip 1em plus 0.5em minus 0.4em\relax {ISCA}, 2022, pp. 296--300.

\bibitem{DBLP:conf/nips/0004LW023}
T.~Liu, K.~A. Lee, Q.~Wang, and H.~Li, ``Disentangling voice and content with self-supervision for speaker recognition,'' in \emph{NeurIPS 2023, New Orleans, LA, USA, December 10 - 16, 2023}, 2023.

\bibitem{DBLP:journals/taslp/LiuLWL24}
------, ``Golden gemini is all you need: Finding the sweet spots for speaker verification,'' \emph{{IEEE} {ACM} Trans. Audio Speech Lang. Process.}, vol.~32, pp. 2324--2337, 2024.

\bibitem{DBLP:conf/cvpr/HuSS18}
J.~Hu, L.~Shen, and G.~Sun, ``Squeeze-and-excitation networks,'' in \emph{{CVPR} 2018, Salt Lake City, UT, USA, June 18-22, 2018}.\hskip 1em plus 0.5em minus 0.4em\relax Computer Vision Foundation / {IEEE} Computer Society, 2018, pp. 7132--7141.

\bibitem{DBLP:conf/cvpr/HeZRS16}
K.~He, X.~Zhang, S.~Ren, and J.~Sun, ``Deep residual learning for image recognition,'' in \emph{{CVPR} 2016, Las Vegas, NV, USA, June 27-30, 2016}.\hskip 1em plus 0.5em minus 0.4em\relax {IEEE} Computer Society, 2016, pp. 770--778.

\bibitem{DBLP:journals/pami/GaoCZZYT21}
S.~Gao, M.~Cheng, K.~Zhao, X.~Zhang, M.~Yang, and P.~H.~S. Torr, ``Res2net: {A} new multi-scale backbone architecture,'' \emph{{IEEE} Trans. Pattern Anal. Mach. Intell.}, vol.~43, no.~2, pp. 652--662, 2021.

\bibitem{NagraniCZ17}
A.~Nagrani, J.~S. Chung, and A.~Zisserman, ``Voxceleb: {A} large-scale speaker identification dataset,'' in \emph{Interspeech}.\hskip 1em plus 0.5em minus 0.4em\relax {ISCA}, 2017, pp. 2616--2620.

\bibitem{ChungNZ18}
J.~S. Chung, A.~Nagrani, and A.~Zisserman, ``Voxceleb2: Deep speaker recognition,'' in \emph{Interspeech}.\hskip 1em plus 0.5em minus 0.4em\relax {ISCA}, 2018, pp. 1086--1090.

\bibitem{DBLP:journals/corr/abs-2306-15354}
\BIBentryALTinterwordspacing
S.~Zheng, L.~Cheng, Y.~Chen, H.~Wang, and Q.~Chen, ``3d-speaker: {A} large-scale multi-device, multi-distance, and multi-dialect corpus for speech representation disentanglement,'' \emph{CoRR}, vol. abs/2306.15354, 2023. [Online]. Available: \url{https://doi.org/10.48550/arXiv.2306.15354}
\BIBentrySTDinterwordspacing

\bibitem{DBLP:conf/interspeech/JungKCJK20}
Y.~Jung, S.~M. Kye, Y.~Choi, M.~Jung, and H.~Kim, ``Improving multi-scale aggregation using feature pyramid module for robust speaker verification of variable-duration utterances,'' in \emph{Interspeech 2020, Shanghai, China, 25-29 October 2020}.\hskip 1em plus 0.5em minus 0.4em\relax {ISCA}, 2020, pp. 1501--1505.

\bibitem{DBLP:conf/icassp/KimSHY22}
J.~Kim, H.~Shim, J.~Heo, and H.~Yu, ``Rawnext: Speaker verification system for variable-duration utterances with deep layer aggregation and extended dynamic scaling policies,'' in \emph{{ICASSP} 2022, Virtual and Singapore, 23-27 May 2022}.\hskip 1em plus 0.5em minus 0.4em\relax {IEEE}, 2022, pp. 7647--7651.

\bibitem{DBLP:conf/slt/MunJHK22}
S.~H. Mun, J.~Jung, M.~H. Han, and N.~S. Kim, ``Frequency and multi-scale selective kernel attention for speaker verification,'' in \emph{{IEEE} Spoken Language Technology Workshop, {SLT} 2022, Doha, Qatar, January 9-12, 2023}.\hskip 1em plus 0.5em minus 0.4em\relax {IEEE}, 2022, pp. 548--554.

\bibitem{DBLP:journals/corr/abs-2305-12838}
\BIBentryALTinterwordspacing
Y.~Chen, S.~Zheng, H.~Wang, L.~Cheng, Q.~Chen, and J.~Qi, ``An enhanced res2net with local and global feature fusion for speaker verification,'' \emph{CoRR}, vol. abs/2305.12838, 2023. [Online]. Available: \url{https://doi.org/10.48550/arXiv.2305.12838}
\BIBentrySTDinterwordspacing

\bibitem{DBLP:conf/icml/IoffeS15}
S.~Ioffe and C.~Szegedy, ``Batch normalization: Accelerating deep network training by reducing internal covariate shift,'' in \emph{{ICML} 2015}, ser. {JMLR} Workshop and Conference Proceedings, vol.~37.\hskip 1em plus 0.5em minus 0.4em\relax JMLR.org, 2015, pp. 448--456.

\bibitem{DBLP:journals/taslp/CaiCZL20}
W.~Cai, J.~Chen, J.~Zhang, and M.~Li, ``On-the-fly data loader and utterance-level aggregation for speaker and language recognition,'' \emph{{IEEE} {ACM} Trans. Audio Speech Lang. Process.}, vol.~28, pp. 1038--1051, 2020.

\bibitem{DBLP:journals/corr/SnyderCP15}
D.~Snyder, G.~Chen, and D.~Povey, ``{MUSAN:} {A} music, speech, and noise corpus,'' \emph{CoRR}, vol. abs/1510.08484, 2015.

\bibitem{DBLP:conf/icassp/KoPPSK17}
T.~Ko, V.~Peddinti, D.~Povey, M.~L. Seltzer, and S.~Khudanpur, ``A study on data augmentation of reverberant speech for robust speech recognition,'' in \emph{{ICASSP} 2017}.\hskip 1em plus 0.5em minus 0.4em\relax {IEEE}, 2017, pp. 5220--5224.

\bibitem{wang2020dku}
W.~Wang, D.~Cai, X.~Qin, and M.~Li, ``The dku-dukeece systems for voxceleb speaker recognition challenge 2020,'' \emph{arXiv preprint arXiv:2010.12731}, 2020.

\bibitem{DBLP:conf/icassp/ThienpondtDD21}
J.~Thienpondt, B.~Desplanques, and K.~Demuynck, ``The idlab voxsrc-20 submission: Large margin fine-tuning and quality-aware score calibration in {DNN} based speaker verification,'' in \emph{{ICASSP} 2021}.\hskip 1em plus 0.5em minus 0.4em\relax {IEEE}, 2021, pp. 5814--5818.

\bibitem{chen20243d}
Y.~Chen, S.~Zheng, H.~Wang, L.~Cheng, T.~Zhu, C.~Song, R.~Huang, Z.~Ma, Q.~Chen, S.~Zhang \emph{et~al.}, ``3d-speaker-toolkit: An open source toolkit for multi-modal speaker verification and diarization,'' \emph{arXiv preprint arXiv:2403.19971}, 2024.

\bibitem{DBLP:conf/interspeech/ZhangLWZH0LM22}
Y.~Zhang, Z.~Lv, H.~Wu, and et~al., ``Mfa-conformer: Multi-scale feature aggregation conformer for automatic speaker verification,'' in \emph{Interspeech 2022, Incheon, Korea, 18-22 September 2022}.\hskip 1em plus 0.5em minus 0.4em\relax {ISCA}, 2022, pp. 306--310.

\bibitem{van2008visualizing}
L.~Van~der Maaten and G.~Hinton, ``Visualizing data using t-sne.'' \emph{Journal of machine learning research}, vol.~9, no.~11, 2008.

\bibitem{DBLP:journals/taslp/GuGZ23}
B.~Gu, W.~Guo, and J.~Zhang, ``Memory storable network based feature aggregation for speaker representation learning,'' \emph{{IEEE} {ACM} Trans. Audio Speech Lang. Process.}, vol.~31, pp. 643--655, 2023.

\bibitem{DBLP:conf/icassp/LiuDLL22}
T.~Liu, R.~K. Das, K.~A. Lee, and H.~Li, ``{MFA:} {TDNN} with multi-scale frequency-channel attention for text-independent speaker verification with short utterances,'' in \emph{{ICASSP} 2022, Virtual and Singapore, 23-27 May 2022}.\hskip 1em plus 0.5em minus 0.4em\relax {IEEE}, 2022, pp. 7517--7521.

\bibitem{moco}
W.~Xia, C.~Zhang, C.~Weng, M.~Yu, and D.~Yu, ``Self-supervised text-independent speaker verification using prototypical momentum contrastive learning,'' in \emph{ICASSP 2021}, 2021, pp. 6723--6727.

\bibitem{rdino}
Y.~Chen, S.~Zheng, H.~Wang, L.~Cheng, and Q.~Chen, ``Pushing the limits of self-supervised speaker verification using regularized distillation framework,'' in \emph{ICASSP 2023}, 2023, pp. 1--5.

\end{thebibliography}

\end{document}